\documentclass[apjl]{emulateapj} 

\usepackage{epsfig}
\usepackage{amsmath}
\usepackage{amssymb}
\usepackage{natbib}
\usepackage{threeparttable} 
\usepackage{booktabs}

\usepackage{epstopdf}

\newcommand{\chan}{\textit{Chandra}}

\newcommand{\swift}{\textit{Swift}}

\newcommand{\Msun}{\mathrm{M}_{\odot}}
\newcommand{\lum}{\mathrm{erg~s}^{-1}}
\newcommand{\flux}{\mathrm{erg~cm}^{-2}~\mathrm{s}^{-1}}

\newcommand{\cnts}{\mathrm{counts~s}^{-1}}

\newcommand{\nh}{\mathrm{cm}^{-2}}
\newcommand{\dens}{\mathrm{cm}^{-3}}
\newcommand{\dist}{(D/8~\mathrm{kpc})^2}

\newcommand{\kms}{\mathrm{km~s}^{-1}}
\newcommand{\gmc}{GM/c^2}

\newcommand{\source}{GRO~J1744--28}
\newcommand{\igr}{IGR J17480--2446}
\newcommand{\rapidburster}{MXB 1730--2335}
\newcommand{\dipper}{4U~1323--62}
\newcommand{\gxns}{GX~340+0}
\newcommand{\exo}{EXO 0748--676}

\slugcomment{Accepted for publication in ApJ Letters}

\shorttitle{\source}
\shortauthors{Degenaar et al.}

\begin{document}

\title{High-Resolution X-Ray Spectroscopy of the Bursting Pulsar GRO J1744--28}

\author{N. Degenaar$^{1,}$\altaffilmark{6}, J.M. Miller$^{1}$, F.A. Harrison$^{2}$, J.A. Kennea$^{3}$, C. Kouveliotou$^{4}$, G. Younes$^{5}$ 
}
\affil{$^1$Department of Astronomy, University of Michigan, 1085 South University Avenue, Ann Arbor, MI  48109, USA; degenaar@umich.edu\\
$^2$Cahill Center for Astronomy and Astrophysics, California Institute of Technology, Pasadena, CA, 91125 USA\\
$^3$Department of Astronomy and Astrophysics, 525 Davey Lab, Pennsylvania State University, University Park, PA 16802, USA\\
$^4$Space Science Office, ZP12, NASA Marshall Space Flight Center, Huntsville, AL 35812, USA\\
$^5$Universities Space Research Association, 6767 Old Madison Pike, Suite 450, Huntsville, AL 35806, USA
}
\altaffiltext{6}{Hubble Fellow}

\begin{abstract}
The bursting pulsar \source\ is a Galactic low-mass X-ray binary that distinguishes itself by displaying type-II X-ray bursts: brief, bright flashes of X-ray emission that likely arise from spasmodic accretion. Combined with its coherent 2.1 Hz X-ray pulsations and relatively high estimated magnetic field, it is a particularly interesting source to study the physics of accretion flows around neutron stars. Here we report on \chan/HETG  observations obtained near the peak of its bright 2014 accretion outburst. Spectral analysis suggests the presence of a broad iron emission line centered at $E_{\mathrm{l}} \simeq6.7$~keV. Fits with a disk reflection model yield an inclination angle of $i \simeq52^{\circ}$ and an inner disk radius of $R_{\mathrm{in}} \simeq 85~\gmc$, which is much further out than typically found for neutron star low-mass X-ray binaries. Assuming that the disk is truncated at the magnetospheric radius of the neutron star, we estimate a magnetic field strength of $B\simeq (2-6)\times10^{10}$~G. Furthermore, we identify an absorption feature near $\simeq6.85$~keV that could correspond to blue-shifted Fe~{\sc xxv} and point to a fast disk wind with an outflow velocity of $v_{\mathrm{out}} \simeq (7.5-8.2)\times10^{3}~\kms$ ($\simeq0.025c-0.027c$). If the covering fraction and filling factor are large, this wind could be energetically important and perhaps account for the fact that the companion star lost significant mass while the magnetic field of the neutron star remained strong.
\end{abstract}


\section{Introduction}
\source, also known as ``the bursting pulsar'', is a transient X-ray binary that was discovered in 1995 \citep[][]{fishman1995,kouveliotou1996}. It displayed two major accretion outbursts, in 1995 and 1996, reaching a luminosity of $L_{\mathrm{X}}$$\simeq$$10^{37}-10^{38}~\dist~\lum$ \citep[e.g.,][]{giles1996,woods1999}. The compact primary is an X-ray pulsar that spins at 2.1~Hz and is thought to have a high magnetic field on the order of $B\simeq10^{11}$~G \citep[e.g.,][]{cui1997,rappaport1997}. X-ray pulse arrival studies revealed an orbital period of 11.8~d, and pointed to a low-mass donor star \citep[][]{finger1996}. Indeed, identification of the near-infrared counterpart suggests that the companion is most likely a late G or K{\sc iii} star \citep[][]{gosling2007,masetti2014_2}. \source\ therefore classifies as a low-mass X-ray binary (LMXB).  

A puzzling property of \source\ is its very low mass-function: it implies that unless the binary is viewed nearly pole-on, which is a priori highly unlikely, the companion star must be of particularly low mass \citep[][]{finger1996}. Evolutionary considerations then suggest that significant mass-transfer must have occurred over the lifetime of the binary \citep[e.g.,][]{bildsten1997_groj1744,rappaport1997,vanparadijs1997}. However, this is difficult to reconcile with the X-ray properties of \source: neutron stars that accreted a large amount of mass typically have much lower inferred magnetic fields ($B\lesssim10^{9}~G$) and much higher spin frequencies \citep[few hundred Hz; e.g.,][]{patruno2012_review}. 

Another remarkable property of \source\ are its bright type-II X-ray bursts \citep[][]{kouveliotou1996}. The hard X-ray spectrum, rapid recurrence time, and radiated energy output suggests that these events result from spasmodic accretion \citep[e.g.,][]{kouveliotou1996,lewin1996,bildsten1997_groj1744}. Only two of the $>$100 known Galactic neutron star LMXBs display type-II bursts \citep[the other being ``the rapid burster'' \rapidburster, e.g.,][]{lewin1996}. 

After remaining dormant for nearly 20 yr, \source\ entered a new bright outburst in 2014 January \citep[e.g.,][]{kennea2014,negoro2014}. Because of its X-ray pulsations and type-II bursts, the source is of special interest to study accretion onto neutron stars and the role of the stellar magnetic field. High-resolution X-ray spectroscopy is particularly promising in this respect: (narrow) absorption features may reveal the presence of energetically-important disk winds \citep[e.g.,][]{miller2006,miller2008,neilsen2009,king2012}, whereas (broad) reflected emission lines can be used to probe the inner radial extent of the accretion disk that may be truncated by the magnetic field of the neutron star \citep[e.g.,][]{bhattacharyya2007,cackett2009_iron,papitto2009,miller2011,miller2013_serx1}. In this Letter we report on \chan/HETG observations obtained near the peak of the 2014 outburst of \source.

\section{\chan/HETG Observations}~\label{sec:obs}
\begin{figure}
 \begin{center}
\includegraphics[width=8.0cm]{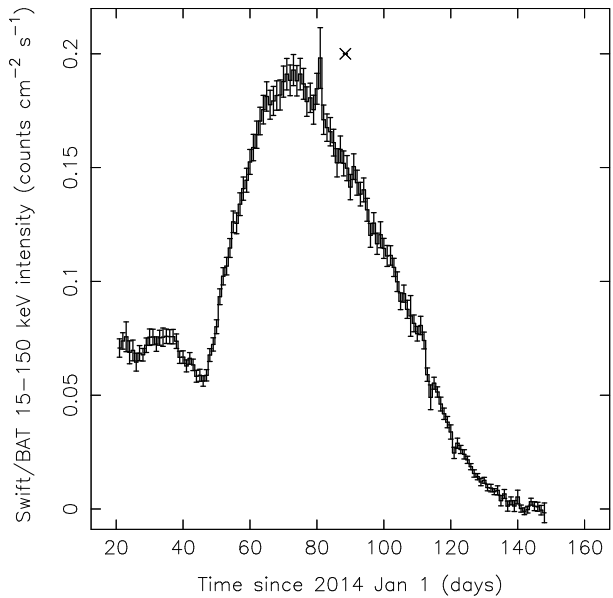}
    \end{center}
\caption[]{{\swift/BAT daily X-ray light curve of the 2014 outburst (15--150 keV), obtained through the transient monitoring program \citep[][]{krimm2013}. The time of our \chan/HETG observation is marked by the cross.\\}}
 \label{fig:bat}
\end{figure}

\source\ was observed with \chan\ on 2014 March 29 (starting at 18:54~{\sc ut}, obsID 16605) and 31 (starting at 00:30~{\sc ut}, obsID 16606), each for a duration of $\simeq$35~ks. As shown in Figure~\ref{fig:bat}, the \chan\ observations occurred near the peak of the outburst. 
The data were obtained with the High Energy Transmission Grating (HETG), which consists of the Medium Energy Grating (MEG; $0.4-5$~keV) and the High Energy Grating (HEG; $0.8-8$~keV). The incoming light was dispersed onto the ACIS-S array, which was operated in continued-clocking (CC) mode to mitigate the effects of pile-up. A gray window was used over the zeroth-order aim point to limit telemetry saturation and loss of frames, and our target was offset from the nominal aim point to avoid CCD chip gaps from impacting the spectra between 5 and 8 keV \citep[see][]{miller2006_winds}. 

We extracted light curves and spectra from the event lists using standard tools within \textsc{ciao} 4.5. The first-order positive and negative grating data were combined, and spectral response files were created using \textsc{mkgrmf} and \textsc{fullgarf}. Given the high hydrogen column density, we focussed our spectral analysis on the HEG data. Spectral fits, carried out within \textsc{XSpec} 12.8 \citep[][]{xspec}, were restricted to the 3--8 keV energy range to avoid instrumental artifacts seen for highly absorbed, bright sources observed in CC mode.\footnote[7]{\chan\ Proposer's Observatory Guide ver. 16.0, section 5.8.1.} Throughout this work we assume a source distance of $D=8$~kpc. Quoted errors refer to 1$\sigma$ confidence levels.

\section{X-Ray Light Curve}\label{sec:lc}
\begin{figure*}
 \begin{center}
\includegraphics[width=18.0cm]{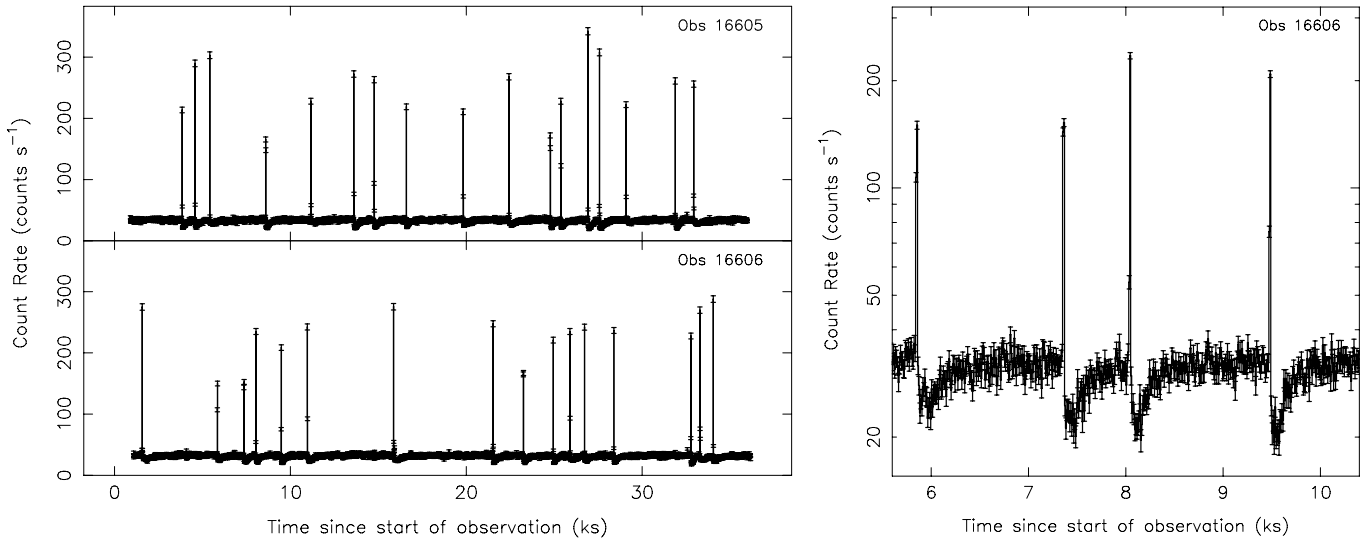}
    \end{center}
\caption[]{{X-ray light curves extracted from the first-order HEG and MEG data at 10~s resolution. Left: Full light curves of the two observations (linear scaling). Right: Zoom of a portion of the light curve of the second observation to highlight the structure of the repetitive bursts (logarithmic scaling).}}
 \label{fig:lc}
\end{figure*}

Figure~\ref{fig:lc} displays the light curves of \source\ during our \chan\ observations. The steady continuum emission is punctuated by several (type-II) bursts that are immediately followed by dips. During the first 35~ks exposure 17 bursts were recorded and during the second 16, implying an average occurrence rate of $\simeq$1.6--1.7~hr$^{-1}$. During the bursts the count rate increases by a factor of $\simeq$4--10 with respect to the steady emission, whereas during the dips it decreases by $\simeq$10\%--50\%. 

\section{Phenomenological Spectral Modeling}\label{sec:spec}
\subsection{Continuum Spectrum}~\label{subsec:cont}
In this work we focus on line features in the X-ray spectra and only provide a basic characterization of the continuum emission. We modeled the combined time-averaged first-order spectra of the two observations using a black body (\textsc{bbodyrad} in \textsc{XSpec}) and a power law (\textsc{pegpwrlw}). To account for interstellar absorption we included the \textsc{tbnew} model using \textsc{wilm} abundances and \textsc{vern} cross-sections \citep[][]{verner1996,wilms2000}. The Fe abundance was allowed to vary to take into account that if the interstellar absorption is high, small deviations from solar composition can have a significant effect on the neutral iron edge near $7$~keV \citep[e.g.,][]{dai2009,cackett2010_iron}. 

Using this continuum model we obtained $N_H=(8.62\pm 0.07)\times10^{22}~\nh$, a black body temperature and emission radius of $kT_{bb}=0.75\pm0.01$~keV and $R_{bb}=14.2\pm0.1$~km (for $D=8$~kpc), and a power-law index and normalization of $\Gamma=0.99\pm 0.01$ and $F_{\mathrm{X,pow}}=(9.02\pm0.02)\times10^{-9}~\flux$ (2--10 keV). The total unabsorbed 2--10 keV model flux is $F_{\mathrm{X}}=(9.74\pm0.01)\times10^{-9}~\flux$, which translates into a luminosity of $L_{\mathrm{X}}=7.5\times10^{37}~\dist~\lum$. The power-law component accounts for $\simeq$90\% of this 2--10 keV unabsorbed flux. The fit resulted in $\chi_{\nu}^{2}=1.56$ for 1027 degrees of freedom (dof). 

\subsection{Emission and Absorption Lines}~\label{subsec:pheno}
\begin{figure}
 \begin{center}
\includegraphics[width=8.0cm]{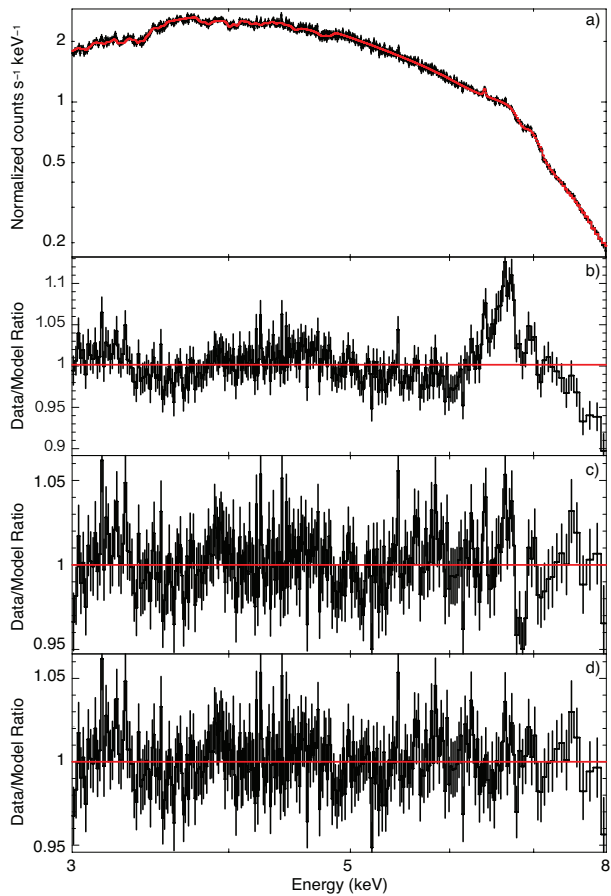}
    \end{center}
\caption[]{{Time-averaged first-order \chan/HEG spectrum. The solid line in the top panel (a) indicates the best fit described in the text. The lower panels indicate the data to model ratio for b) the continuum model, c) the ratio after adding one broad Gaussian emission line near $\simeq$6.7~keV, and d) the ratio of the best fit (i.e., after including a narrow emission line at $\simeq$6.40~keV and two narrow absorption lines at $\simeq$6.85 and 5.11~keV).}}
 \label{fig:specfit}
\end{figure}

Our continuum fits leave significant residuals between 6--7 keV (panel b in Figure~\ref{fig:specfit}). We attempted to model these features with simple Gaussians (\textsc{gauss}). The structure is complex, however, and we did not find a unique solution. An acceptable fit ($\chi_{\nu}^{2}=1.05$ for 1017 dof) can be obtained by using a combination of three emission lines: such features are seen for X-ray pulsars in high-mass X-ray binaries (HMXBs), but not for LMXBs \citep[e.g.,][]{torrejon2010}. This description is explored in a companion paper (Younes et al., in preparation). Here, we focus on a different physical interpretation that is more common for LMXBs and could possibly account for some of the peculiar properties of \source\ (Section~\ref{subsec:implications}).

Adding a first Gaussian significantly improves the fit ($\chi_{\nu}^{2}=1.08$ for 1024 dof; f-test probability $\simeq10^{-85}$), and suggests the presence of a broad emission line. It has a central energy of $E_{\mathrm{l}}=6.71\pm0.02$~keV, and an equivalent width of $EW=117.0\pm 4.8$~eV (see Table~\ref{tab:spec} for detailed properties). The line energy is consistent with ionized (He-like) iron Fe~{\sc xxv} (rest energy $6.70$~keV). Such lines are often seen in LMXBs (Section~\ref{subsec:iron}). Dividing the normalization by its (minus-sign) error suggests that the feature is significant at the $\simeq$25$\sigma$ level of confidence. 

Clear deviations above and below the model fit remain near $\simeq$6.4 and 6.8 keV, respectively (panel c in Figure~\ref{fig:specfit}). The addition of two more Gaussians (one in emission, one in absorption) further improves the fit ($\chi_{\nu}^{2}=1.05$ for 1017 dof; ftest-probability of $\simeq10^{-5}$). Both lines are narrow, with an equivalent width of $EW=2.2\pm0.6$~eV ($E_{\mathrm{l}}=6.392\pm 0.002$~keV) and $EW=-8.7\pm 1.6$~eV ($E_{\mathrm{l}}=6.85^{+0.24}_{-0.13}$~keV) for the emission and absorption feature, respectively (Table~\ref{tab:spec}). The emission line at $\simeq$6.40 keV is compatible with neutral iron (Fe~{\sc i} up to Fe~{\sc x}) and is significant at the $\simeq4\sigma$ level of confidence. The absorption feature at $\simeq$6.85 keV is $\simeq6\sigma$ significant. It could possibly correspond to redshifted Fe~{\sc xxvi} (rest energy 6.97 keV), e.g. arising from matter moving along the magnetic field lines. However, assuming that the velocity shift of $\simeq0.017c$ is a gravitational redshift, the implied radial distance would be $\simeq3.5\times10^{3}~\gmc$, which is much further out than the estimated magnetospheric radius of $\simeq85~\gmc$ (Section~\ref{subsec:broad}). Alternatively, and perhaps more likely, this absorption feature could correspond to blue-shifted Fe~{\sc xxv} and arise from a disk wind, as is often seen in LMXBs (Section~\ref{subsec:winds}). 

An additional absorption feature at $E_{\mathrm{l}}=5.11\pm 0.02$~keV (with an unresolved width; Table~\ref{tab:spec}) appears to be $\simeq$3$\sigma$ significant. Photoionization modeling suggest that producing absorption at this energy would require unusual abundances (e.g., of Mn or Cr) and unusually high velocity shifts. Given the relatively low significance, it is plausible that the feature is spurious. However, we briefly consider the possibility that it is an electron cyclotron resonance line. For the fundamental energy of $E_e = 11.6 (1+z)^{-1} (B/10^{12}~\mathrm{G})$~keV to fall at $5.11$~keV, the magnetic field of the neutron star would have to be of the order of $B\simeq 6 \times10^{11}$~G (for a gravitational redshift of $1+z=1.31$). This is broadly consistent with previous estimates for \source\ \citep[e.g.,][]{finger1996,bildsten1997_groj1744,rappaport1997,cui1997}, but a factor $\simeq$10 higher than indicated by modeling the broad Fe emission line (see Section~\ref{subsec:broad}). The final fit, including this additional absorption line, yielded $\chi_{\nu}^{2}=1.04$ for 1015 dof and is shown in Figure~\ref{fig:specfit}.

\subsection{Broad Emission Line Modeling}~\label{subsec:iron}
Broad Fe-emission lines are often seen in black hole \citep[e.g.,][for a review]{miller2007} and neutron star LMXBs \citep[e.g.,][]{bhattacharyya2007,papitto2009,cackett2010_iron,miller2011,miller2013_serx1}. These are thought to result from reflection of the accretion disk that is being illuminated by the central X-ray source. 

We therefore also attempted to model the broad emission feature of \source\ by a physical reflection model. Using \textsc{diskline} \citep[][]{fabian1989} we obtain a good fit ($\chi_{\nu}^{2}=1.05$ for 1014 dof). We find an inner accretion disk radius of $R_{\mathrm{in}}=85.0\pm10.9~GM/c^2$ and a disk inclination of $i = 51.8{^\circ}\pm4.0{^\circ}$. For the emissivity index we obtain $\beta= -2.54\pm 0.05$, whereas the outer disk radius was kept fixed at $R_{\mathrm{out}}=10^4~GM/c^2$.

\subsection{Bursts, Dips and Steady Emission}~\label{subsec:bursts}
We made count rate cuts to investigate possible spectral changes between the bursts, dips and steady emission seen in the X-ray light curves (Figure~\ref{fig:lc}). For this purpose we extracted a burst spectrum from data with $>$50~$\cnts$, a dip spectrum for $<$28$~\cnts$, and a steady emission spectrum for 30--40$~\cnts$. 

Due to the lower statistics of the time-sliced spectra it was not possible to reliably determine any changes in the emission/absorption features. For reference, we report that the 2--10 keV unabsorbed continuum fluxes of the different intervals are $F_{\mathrm{X}}= (5.38\pm 0.04)\times10^{-8}$ (bursts), $(7.32 \pm 0.03)\times10^{-9}$ (dips), and $(9.93\pm 0.03)\times10^{-9}~\flux$ (steady). By multiplying the obtained average fluxes with the exposure time of the spectra, we find that the ratio of the integrated energy of the steady/burst emission is $\alpha \simeq 17$.

\section{Photo-ionization Modeling}~\label{sec:xstar}
To construct a physical picture of the absorbing and emitting gas in \source, we created a grid of self-consistent photoionization models with \textsc{xstar} \citep[][]{kallman2001}. We modeled the ionizing flux as a $\Gamma=1$ power law with an estimated bolometric luminosity of $L=10^{38}~\lum$. We assumed a covering factor of $\Omega/4\pi=0.67$, a maximum density of $n=10^{12}~\dens$, a turbulent velocity of $200~\kms$, and solar abundances. 

\begin{figure}
 \begin{center}
\includegraphics[width=8.0cm]{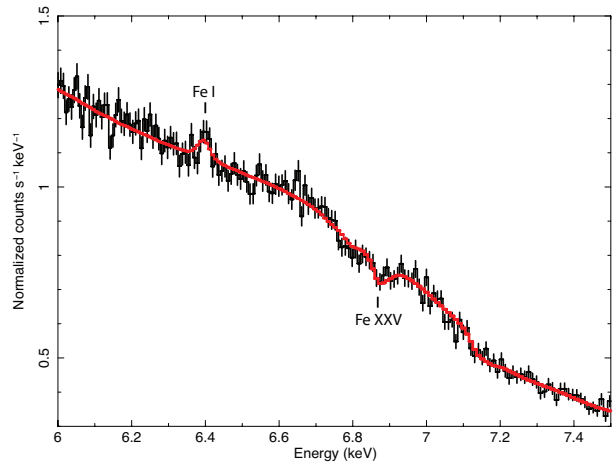}
    \end{center}
\caption[]{{Time-averaged first-order \chan/HEG spectrum together with a fit using an \textsc{xstar} grid to model the narrow emission feature near $\simeq$6.40~keV (neutral iron) and the absorption feature near $\simeq$6.85~keV (Fe~{\sc xxv}). }}
 \label{fig:xstarfit}
\end{figure}

Using this grid of models, we can fit the absorption line near $\simeq6.85$~keV with a density of $N=3.7^{+0.6}_{-0.3}\times10^{21}~\nh$, an ionization parameter of $\log \xi = 3.9\pm0.1$, and a blue shift of $v_{\mathrm{out}}=7868\pm313~\kms$ (Figure~\ref{fig:xstarfit}). Given that $\xi = L / nr^{2}$ and $N = fnr$ (where $f$ is the filling factor), the radial distance of the absorbing medium from the ionizing X-ray source can be estimated as $r=f L/N\xi $. For a uniform filling factor (i.e., $f=1$) this gives $r\simeq 10^{12}$~cm, which is comparable to the binary separation \citep[$\simeq$25~$R_{\odot}$; e.g.,][]{finger1996}. We can regard this as an upper limit, since the actual density of the absorbing material (i.e., the filling factor) is not known. The radial distance is thus consistent with the (outer) accretion disk. 

Using the same grid of \textsc{xstar} models we found that the narrow emission feature near $\simeq$6.40~keV would require a higher density and a lower ionization level ($N\gtrsim10^{23}~\nh$ and $\log \xi \lesssim 2.2$; Figure~\ref{fig:xstarfit}). The fit suggests a modest redshift of $v_{\mathrm{in}} = 690^{+120}_{-570}~\kms$, although fixing the redshift to zero gives an equally good fit. The implied radial distance of $r=f L/N\xi \simeq 2\times 10^{12}$~cm (for $f=1$) is similar to the binary separation. This may suggest that the narrow emission line originates in the wind of the companion star or perhaps the impact point where the matter stream hits the accretion disk. A systematic \chan\ study of narrow Fe~{\sc i} emission lines of LMXBs/HMXBs favored an origin in the (stellar wind of the) companion star \citep[][]{torrejon2010}.

\begin{table*}
\begin{center}
\caption{X-Ray Spectral Parameters for the Summed HEG Data \label{tab:spec}}
\begin{tabular*}{0.98\textwidth}{@{\extracolsep{\fill}}llcc}
\hline
\hline
Model Component & Parameter (Unit) &  \multicolumn{2}{c}{Model for Broad Fe Line}  \\
\hline
 &  & Gaussian  & Diskline \\
TBNEW  & $N_H~(10^{22}~\nh)$ & $10.95 \pm 0.15$ & $10.91\pm0.17$  \\	
  & Fe abundance & $1.03\pm 0.05$ & $0.94\pm 0.05$  \\	
 &  &  &  \\
BBODYRAD  & $kT_{\mathrm{bb}}$~(keV) &   $0.72\pm 0.01$ &   $0.72\pm 0.01$    \\  
 & $R_{\mathrm{bb}}$~(D/8.0 kpc km)  &  $23.56\pm 0.02$ &  $22.82^{+0.02}_{-0.28}$    \\  
 &  &  &  \\
PEGPWRLW  & $\Gamma$~(keV) &  $0.99 \pm 0.01$ &  $1.00\pm 0.01$   \\  
 & $F_{\mathrm{X,pow}}$~($10^{-9}~\flux$)  &  $8.72\pm 0.02$ &  $8.75\pm 0.07$   \\  	
 &  &  &  \\
DISKLINE & $E_{\mathrm{l}}$ (keV)  & \nodata & $6.69\pm 0.02$   \\	
 & $R_{\mathrm{in}}$ ($GM/c^2$) &  \nodata &  $85.0\pm10.9$   \\	    
  & $i$ ($^{\circ}$) &  \nodata &  $51.8\pm 4.0$   \\	
  & $\beta$  &  \nodata &  $-2.54\pm 0.05$   \\	
 & $N_{\mathrm{l}}$ ($10^{-2}$~ph~cm$^{-2}$~s$^{-1}$) &  \nodata &  $1.10\pm 0.04$  \\
    & $EW$ (eV) &   \nodata & $106.1\pm 3.9$  \\	
 &  &  &  \\
GAUSSIAN1: Broad emission (Fe \textsc{xxv}) & $E_{\mathrm{l}}$ (keV) & $6.71\pm 0.02$ & \nodata  \\	
 & $\sigma$ (keV) &  $0.29\pm 0.02$ &  \nodata   \\		  
 & $N_{\mathrm{l}}$ ($10^{-2}$~ph~cm$^{-2}$~s$^{-1}$) &  $1.23\pm 0.05$ &  \nodata   \\
   & $EW$ (eV) &  $117.0\pm 4.8$ &  \nodata   \\	
 &  &  &  \\
GAUSSIAN2: Narrow emission (neutral Fe) & $E_{\mathrm{l}}$ (keV) &  $6.392\pm 0.002$ &   $6.400\pm 0.005$  \\	
  & $\sigma$ (keV) &  $0.01$ fix &  $0.034^{+0.054}_{-0.029}$    \\	
  & $N_{\mathrm{l}}$ ($10^{-4}$~ph~cm$^{-2}$~s$^{-1}$) &  $2.78 \pm 0.76$ & $3.48\pm 0.77$   \\
    & $EW$ (eV) &  $2.2\pm0.6$  &  $2.9\pm 0.6$   \\  	
 &  &  &  \\
GAUSSIAN3: Absorption (blue-shifted Fe \textsc{xxv})  & $E_{\mathrm{l}}$ (keV) &   $6.85^{+0.24}_{-0.13}$ &  $6.85^{+0.24}_{-0.13}$  \\	
 & $\sigma$ (keV) &  $0.05\pm 0.02$ &  $0.05\pm 0.02$   \\	
 & $N_{\mathrm{l}}$ ($10^{-4}$~ph~cm$^{-2}$~s$^{-1}$) &  $-9.9\pm 1.8$ &  $-9.9\pm 1.8$   \\	
  & $EW$ (eV) &  $-8.7\pm 1.6$ &  $-8.7\pm 2.6$   \\	
 &  &  &  \\
GAUSSIAN4: Absorption (unidentified) & $E_{\mathrm{l}}$ (keV) &   $5.11\pm 0.02$ &   $5.11\pm 0.02$  \\	
  & $\sigma$ (keV) &  $0.01$ fix &  $0.01$ fix \\	
  & $N_{\mathrm{l}}$ ($10^{-4}$~ph~cm$^{-2}$~s$^{-1}$) &  $-2.78\pm 0.84$ &  $-2.88\pm 0.84$    \\	
  & EW (eV) &  $-1.8\pm0.5$ &  $-1.9\pm0.5$  \\
 & 	$\chi_{\nu}^2$/dof & 1.04/1015& 1.05/1016   \\
\hline
\end{tabular*}
\tablenotes{
{\bf Notes.} The two right columns represent fits with different descriptions for the broad iron line (Gaussian or disk line; see text). Quoted errors represent a $1\sigma$ confidence level. The Fe abundance for the absorption model (\textsc{tbnew}) is given with respect to solar composition. The \textsc{pegpwrlw} normalization, $F_{\mathrm{X,pow}}$, represents the unabsorbed power-law flux in the 2--10 keV range. \\
}
\end{center}
\end{table*}

\section{Discussion}\label{sec:discussion}
\subsection{Broad Iron Emission Feature: Disk Reflection?}~\label{subsec:broad}
\chan/HEG spectroscopy of \source\ during its 2014 accretion outburst revealed a broad emission line near $6.7$~keV, as was seen during previous outbursts \citep[e.g.,][]{nishiuchi1999}. Fits with a relativistic disk model yield an inner disk radius of $R_{\mathrm{in}}\simeq85~GM/c^2$, which is much further out than typically found for neutron star LMXBs \citep[$\simeq6-15~GM/c^2$;][]{cackett2010_iron}. This larger truncation radius is plausibly due to the higher magnetic field of \source. 

Assuming that the inner disk is truncated at the magnetospheric radius, we can use equation 1 of \citet{cackett2009_iron} to estimate the magnetic field strength of the neutron star. For a bolometric flux of $F_{\mathrm{X}}\simeq2\times10^{-8}~\flux$ (estimated to be a factor of $\simeq2$ higher than measured in the 2--10 keV band), $D$$=$8~kpc, $M$$=$1.4$\Msun$, $R_{\mathrm{in}}$$=$85~$GM/c^2$, and making the same assumptions regarding geometry and the accretion efficiency, we obtain $B\simeq$$ (2-6)\times10^{10}$~G. This is indeed much higher than the magnetic field typically assumed for neutron star LMXBs ($B\lesssim 10^{9}$~G). 

This inferred magnetic field is consistent with estimates based on the pulsar spin up \citep[$B\lesssim3\times10^{11}$~G;][]{bildsten1997_groj1744}, but a factor of a few lower than suggested by evolutionary considerations \citep[$B\simeq(2-7)\times10^{11}$~G;][]{rappaport1997} and the possible onset of a propellor phase during the outburst decay \citep[$B\simeq2\times10^{11}$~G;][]{cui1997}. Therefore, it was previously proposed that the reflecting material may be located inside the magnetic radius \citep[e.g., matter that leaks through the magnetospheric barrier or is channeled along the field lines;][]{nishiuchi1999}. Alternatively, it could imply that Compton scattering contributes to broadening the Fe line \citep[e.g.,][]{ross2007}, although this is likely to be less important than dynamical broadening \citep[e.g.,][]{miller2002,cackett2013_iron}.

\subsection{Iron Absorption Feature: A Fast Disk Wind?}~\label{subsec:winds}
Our spectral analysis suggests the possible presence of an absorption feature near $\simeq$6.85~keV. Fits with an \textsc{xstar} grid suggests that this line could be blue-shifted Fe~{\sc xxv}, which would indicate a very fast outflow of $v_{\mathrm{out}}\simeq(7.5-8.2)\times10^{3}~\kms$ ($\simeq0.025-0.027c$) at a radial distance of $r \lesssim10^{12}$~cm, consistent with the outer accretion disk. Since the outflow speed is significantly higher than the typical terminal velocity of stellar winds, it seems likely that the absorption occurs in a disk wind. 

We can estimate the mechanical luminosity in this wind as $L_{\mathrm{w}}=0.5 f m_p \Omega v_{\mathrm{out}}^3 L/\xi$, where $m_p$ is the proton mass \citep[][]{miller2011}. Based on our \textsc{xstar} fits ($\Omega/4\pi=0.67$, $L=10^{38}~\lum$, $v_{\mathrm{out}}=7868~\kms$, and $\log \xi = 3.9$) we then find $L_{\mathrm{w}}\simeq 4 \times 10^{37}~\lum$, or $L_{\mathrm{w}}/L_{\mathrm{acc}} \simeq 0.4$ for a uniform filling factor ($f=1$). This implies that if the covering fraction and filling factor are large, the wind in \source\ may carry away a significant amount of the mass that is transferred from the companion star.

Although disk winds have now been detected in several black hole LMXBs \citep[e.g.,][for a recent overview]{ponti2012_winds}, it is not yet established whether similarly energetic outflows can also form in neutron star systems. Absorption features have been seen in several high-inclination sources (eclipsers and dippers). However, these do not show significant blue shifts, indicating that the absorption might occur in the disk atmosphere and does not necessarily require an outflow \citep[e.g., \gxns, \dipper, and \exo;][]{boirin2005,cackett2010_iron,ponti2014_exo0748}. The best cases for disk winds in neutron star LMXBs are Cir X-1 \citep[e.g.,][]{brandt2000}, GX 13+1 \citep[e.g.,][]{ueda2004}, and \igr\ \citep[][]{miller2011}. The estimated outflow velocities in these three sources is $v_{\mathrm{out}}\simeq(0.4-3.0)\times10^{3}~\kms$. 

It is of note that both \source\ and \igr\ are X-ray pulsars (spinning at 2.1 and 11 Hz, respectively), that GX 13+1 shows similar behavior as `Z-sources' \citep[e.g.,][]{homan1998}, and that Cir X-1 is an exceptionally young LMXB \citep[][]{heinz2013}. This suggests that their magnetic fields may be stronger than the majority of (non-pulsating) LMXBs, which could perhaps play a role in the production of winds \citep[][]{miller2011}. Indeed, a disk wind with a velocity of $v_{\mathrm{out}}\simeq(1.0-1.5)\times10^{3}~\kms$ was also found in the HMXB X-ray pulsar 1A 0535+262 \citep[][]{reynolds2010}. On the contrary, jet formation may be suppressed in neutron stars with higher magnetic fields \citep[e.g.,][]{massi2008,migliari2012}.

\subsection{Implications for the Binary Parameters}~\label{subsec:implications}
It was already realized that the very low mass-function of \source, $f=(M_{\mathrm{c}} \sin i)^3/{(M_{\mathrm{c}}+M_{\mathrm{NS}})^2}=1.36\times10^{-4}$, implies that if the viewing geometry is not nearly pole-on, the donor must be of very low mass \citep[][]{finger1996,vanparadijs1997,rappaport1997}. If the broad emission line indicated by our spectral fits is a disk reflection feature, then the inclination of $i\simeq52^{\circ}$ suggests a companion mass of $M_{\mathrm{c}}\simeq 0.07~\Msun$ for $M_{\mathrm{NS}}=1.4~\Msun$. This is well below the mass expected for a late G/K giant companion star \citep[note that mass-loss can leave the spectral type unaffected;][]{gosling2007}, and hence suggests that a significant amount of mass may have been transferred from the donor star.

To reconcile this with the high magnetic field of \source\ (which would be expected to have decayed significantly if a large amount of mass was accreted), it was proposed that most of the mass-transfer occurred before the neutron star formed through accretion-induced collapse of a white dwarf \citep[][]{vanparadijs1997}. However, the neutron star should still have accreted a significant amount of mass to reduce the companion to the present-day value \citep[e.g.,][]{rappaport1997}. If a fast disk wind is present in \source, it could possibly help lift this discrepancy as it may allow for a significant amount of mass to have been lost from the donor without accreting onto the neutron star.

\acknowledgments
We thank Harvey Tananbaum, Belinda Wilkes, and Andrea Prestwich for approving and executing this DDT observation. N.D. is supported by NASA through Hubble Postdoctoral Fellowship grant number HST-HF-51287.01-A from the Space Telescope Science Institute, which is operated by the Association of Universities for Research in Astronomy, Incorporated, under NASA contract NAS5-26555. N.D. thanks Ed Cackett, Felix F{\"u}rst and Rudy Wijnands for valuable discussions. We extend our thanks to the anonymous referee.

{\it Facility:} \facility{CXO (HETG)}

\end{document}